\documentclass[12pt,preprint]{aastex}
\begin{document}
   \title{Main-sequence stellar eruption model for V838 Mon}

\author{Noam Soker\altaffilmark{1,2}
           and
     Romuald Tylenda\altaffilmark{1} }

\altaffiltext{1}{Copernicus Astronomical Center, Department for Astrophysics, 
              Rabia\'nska 8, 87-100 Toru\'n, Poland; 
              tylenda@ncac.torun.pl}
 
\altaffiltext{2}{Department of Physics, Oranim, Tivon 36006, Israel;
             soker@physics.technion.ac.il}

\begin{abstract}

We propose that the energy source of the outburst of V838 Mon 
and similar objects is an accretion event, i.e., gravitational 
energy rather than thermonuclear runaway.
We show that the merger of two main sequence stars, of masses 
$M_1 \simeq 1.5 M_\odot$, and $M_2 \simeq 0.1-0.5 M_\odot$
can account for the luminosity, large radius, and low effective 
temperture of V838 Mon and similar objects. Subsequent cooling 
and gravitational contraction lead such objects to move 
along the Hayashi limit, as observed.
By varying the masses and types of the merging stars, and
by considering slowly expanding, rather than hydrostatic,
envelopes, this model can account for a large range in 
luminosities and radii of such outburst events. 

\end{abstract}

%

\section{Introduction}

V838 Mon (Munari et al. 2002, hereafter \cite{mhk}; 
Kimeswenger et~al. \cite{kls}), 
V4332 Sgr (Martini et al. \cite{mwt}), and most probably M31~RV 
(M31~Red~Variable -- Mould et al. \cite{mcg}),
form a peculiar and mysterious group of erupting stellar objects.
They do not fit to any known class of stellar outbursts. 
Just from the light curve they can be classified as slow novae but 
their spectral evolution clearly shows that they are not. 
After developing a K-type giant-supergiant spectrum near maximum they evolve
to lower effective temperature and fade as very late M-type giants or
supergiants. The bolometric fading at a very low effective temperature has
undoubtly been observed for V4332~Sgr (Martini et~al. \cite{mwt}) and M31~RV
(Mould et~al. \cite{mcg}). It seems that also
V838~Mon is evolving in the same direction (e.g. Osiwa{\l}a et~al. \cite{omt}). 
No hot phase, as e.g. nebular stage in classical novae, has been observed in
any of the three objects.
These observational facts almost certainly exclude any kind of thermonuclear event
models, such as runaway models for classical novae (see e.g. that proposed for
M31~RV by Iben \& Tutukov \cite{it})
or a late post-AGB He-shell flash (born-again AGB), which are usually discussed while 
trying to interpret the above variables (\cite{mhk}; Kimesweneger et~al. \cite{kls}). 
The point is that the thermonuclear models after relatively cool phase at
(visual) maximum always evolve to higher and higher effective temperature
before they star fading in luminosity. The reason is that the 
stellar envelope, inflated by the initial thermonuclear outburst, starts 
shrinking well before the nuclear reactions start declining. 

As usually in discussions of nature of a stellar outburst the observational
data on the pre-outburst object are very important. Unfortunantely they are
very scant in our case. Photometric data for V838~Mon in pre-outburst are 
consistent with an F-type main sequence star (\cite{mhk}; 
Kimesweneger et~al. \cite{kls}), althought observational uncertainities are 
important. For V4332~Sgr a K-type star is suggested as a progenitor 
(Martini et~al., \cite{mwt}). 
There are no data on the pre-outburst state of M31~RV.

If a thermonuclear event is excluded as argued above, the next energy source 
that comes to mind is an accretion (or merging) event. 
We therefore investigate in this paper a scenario in which a $\sim$1.5~M$_\odot$ 
(F-type) main sequence star accretes a less massive objects. 
Given the energy
liberated during the observed outbursts it is clear that accretion of at
least a massive planet onto the main sequence star has to be involved. 

While constraining our model we use the results from observations of V838~Mon,
as this is the best observed object from the three (e.g. \cite{mhk},
Kimesweneger et~al. \cite{kls}, Kolev et~al. \cite{kmt},
Osiwa{\l}a et~al. \cite{omt}). 
This star erupted in the begining of 2002, first by $\sim$6~mag. (in V)
and as a K-type star it reached a luminosity of 
$\sim 500(D/1{\rm kpc})^2\mbox{L}_\odot$, and a radius of 
$\sim 55(D/1{\rm kpc})\mbox{R}_\odot$, where $D$ is the distance to V838 Mon.
Next, after a month of a slow decline and cooling it had another brightning
by $\sim$4.5~mag. and reached a luminosity of
$\sim 1.0 \times 10^4 (D/1 {\rm kpc})^2 \mbox{L}_\odot$, and a radius of
$\sim 120 (D/1 {\rm kpc}) \mbox{R}_\odot$.
Subsequent evolution can be described as a more or less horizontal evolution
on the H-R diagram towards lower and lower effective temperatures somewhat
interupted by a minor brightnening a month after the second outburst.
The effective radius has countinuously been increasing and in April reached 
a value of $\sim 400 (D/1 {\rm kpc}) \mbox{R}_\odot$ (\cite{mhk}). 
Numerous spectral lines have been observed to show P-Cyg profiles during the
outburst indicating an outflow with a velocity of 200-500~km\,${\rm s}^{-1}$,
but no mass loss rate estimate is available. Lines of LiI and BaII have 
been observed, but no analysis of the element abundances has been done.
In mid April the star started a rapid decline in
V accompanied by the spectral evolution to late M-types (Osiwa{\l}a et~al.
\cite{omt}). 

A scattered light echo have been detected around V838~Mon
and it was used to drive its distance.
While \cite{mhk} derive a distance of $D = 0.79$ kpc and 
Kimeswenger et~al. (\cite{kls}) derive a distance of $D=0.66$~kpc,
a calcualtion by Bond et al.\ (\cite{bps}; 
H.E. Bond, private communication) 
yields a much larger distance of $D \simeq 3$~kpc. 
Above, and in the rest of the paper, we scale the properties 
of V838 Mon with the distance, such that our model covers this
entire distance scale. 
Indeed, one of the adavantages of our proposed model, is that
it can account for a large range of luminosities, because the 
luminosity is fixed by the type and mass of the merging stars. 
Thus we can also account for M31~RV, which was significantly more luminous 
than the other two objects and reached $\sim 10^6 \mbox{L}_\odot$

Our paper is organized as follows.
In Sect. 2 we estimate the general properties of a main 
sequence star, the primary, which accretes mass from a 
destructed main sequence companion.
These results are used in Sect. 3 to proposed a scenario 
to the eruption of V838 Mon. 
Our discussion and summary are in Sect. 4.

\section{Accreting main sequence stars}

Accreting main sequence stars were found to expand to 
giant dimensions
in several different kind of calculations.
These include the merger of two main sequence stars, 
in studying the formation of blue straggler stars
(e.g. Sills et al. \cite{slb}), mass accretion by a main 
sequence companion inside an envelope of a 
giant (Hjellming \& Taam 1991, hereafter \cite{ht}), 
and an accreting low mass main sequence star 
(Prialnik \& Livio 1985, hereafter \cite{pl}). 

PL85 calculate the outcome of an accreting fully convective 
$0.2 M_\odot$ main sequence star.
When a large fraction of the free fall energy of the accreted 
matter is retained, they find the star to expand.
\cite{pl} parametrized the fraction of energy retain by a 
parameter $\alpha$,
\begin{equation}
U= \alpha  {G M_\ast M_{\rm acc}}/{R_\ast} ,
\end{equation}
where $M_\ast$ and $R_\ast$ are the 
mass and redius of the accreting main sequence star, respectively,
and $M_{\rm acc}$ is the accreted mass to the envelope.
For an orbiting companion that is being destructed and accreted,
the maximum available energy has $\alpha=0.5$.  
Not all this energy will be retain by the accreted matter 
which build the envelope in our model. 
Large fraction of the energy goes to radiaton and mass loss
observed in the eruption of V838 Mon and the other objects.
\cite{pl} calculate the accretion of mass up to 
$\Delta M = 2.5 \times 10^{-3} M_\odot$,
and find that when $\alpha$ is not too small the star expands.
A good fit to their results of $\alpha=0.1$ and an accretion 
rate of $\dot M_{acc} = 10^{-2} M_\odot \mathrm{yr^{-1}}$, 
and when the accreted mass is 
$M_{\rm acc} > 0.5 \times 10^{-3} M_\odot$, 
gives for the radius of the star 
\begin{equation}
\frac {R_\ast}{R_\odot} \simeq 
2.2 \left(\frac {M_{\rm acc}}{10^{-3} M_\odot} \right) + 0.5.
\end{equation} 
Extrapolating this fit to much higher accreted mass and radii,
we find that to obtain a radius of $R_\ast \simeq 50-150 R_\odot$
(characteristic fro the first brightening of V838~Mon),
the accreted mass should be $M_{\rm acc} \simeq 0.02-0.07 M_\odot$.
With higher accretion rate and higher values of $\alpha$, \cite{pl}
find larger radii. 
 
In the calculations of \cite{ht}, a $1.25 M_\odot$ main sequence 
star accretes from a giant up to $M_{\rm acc} = 0.115 M_\odot$, 
at an increasing rate up to 
$\dot M_{\rm acc} \simeq 10^{-2} M_\odot \mathrm{yr^{-1}}$. 
The final radius of the accreting star depends on the entropy
of the accreted gas, being lower for lower-entropy accreted gas.
We take the higher entropy case; 
we later take the low entropy-high density
gas to be accreted into the primary star.  
In this case (\cite{ht}'s model 2c), a good fit to figures 1a and 3a 
of \cite{ht} is
\begin{equation}
\frac {R_{\ast}}{R_\odot} \simeq 1.4 \times
10^{17(M_{\rm acc}/M_\odot)}. 
\end{equation} 
This is a much stronger dependence than that found by \cite{pl}, and cannot
be extrapolated to much larger accreted mass since then the radius
becomes too large. 
However, this formula does demonstrate that a small
accreted mass can cause a large expansion when the star is already
a giant. 
By the last fit, to obtain a radius of $R \simeq 300-1500 R_\odot$
(characteristic for the second, major outburst of V838~Mon), 
the accreted mass should be $M_{\rm acc} \sim 0.14 -0.2M_\odot$.
The strong dependance in the last equation, suggests that with 
more energy supplied, as in the cores merger we discussed later,
less mass can be inflated to large radii. 
Despite the large differences between the two works cited above,
and the very different fits to \cite{pl} and \cite{ht}, the amount of 
accreted mass, required to inflate a giant to 
$R_\ast \simeq 300 R_\odot$ is similar, and amounts
to $M_{\rm acc} \sim 0.15 M_\odot$.
In the proposed scenario, two main sequence stars merge on 
a dynamical scale, following dynamical instability
(Rasio \& Shapiro \cite{rs})
and the accretion rate can be as high as 
$\dot M_{\rm acc} \sim 0.03 M_\odot / 1~{\rm day}=  
10 M_\odot \mathrm{yr^{-1}}$.  
At such a high accretion rate, and with most energy deposited in the
accreted mass, an accreted mass lower than $0.15 M_\odot$ 
can be inflated to giant dimensions. 
Therefore, for our scenario we take the inflated envelope mass to be
$M_{\rm acc} \sim 0.05-0.3 M_\odot$. 

In the calculations cited above the accreted mass was added to the
outer layers of the inflated star. In our scenario, the mass is added
at the base of the inflated envelope, at the surface
of the original primary star. 
A somewhat different treatment of the energy budget is required.
Let the destructed companion be split into low entropy-high density gas 
of mass $M_c$, i.e., the core of the companion, which is accreted and 
sinks into the core of the primary star, down to a radius 
$r_f<R_1$ (where $R_1$ is the initial radius of the accreting primary star)  
and a high entropy mass $M_{\rm acc}$ which goes into the inflated envelope.  
This assumption is based on the numerical calculations of 
Sandquist, Bolte \& Hernquist (\cite{sbh}). 
The total energy of this structure is
\begin{equation}
E_g  \simeq - \frac {G M_1 M_c}{r_f} + E_{\rm env}, 
\end{equation} 
where $E_{\rm env}$ is the gravitational energy of the envelope.
For an envelope having a density profile of $\rho = K r^{-2.5}$
from $R_1$ to $R_\ast$, where $K$ is a constant, which crudely 
fits the envelope found by \cite{pl} for the model cited above, 
the envelope energy has a simple approximated form for 
$R_\ast \gg R_1$, given by 
\begin{equation}
E_{\rm env} \simeq - \frac{ G (M_1+M_c)M_{\rm acc}}{R_1} 
     \left(\frac {R_1}{R_\ast} \right)^{1/2}. 
\end{equation} 
The original energy of the binary system is
\begin{equation}
E_0 \simeq - 0.5 \frac {G M_1 M_2}{a}, 
\end{equation} 
where $a$ is the orbital separation at destruction. 
In addition, an energy $E_2>0$ should be supplied to destruct 
the companion.
The condition to inflate the envelope becomes
\begin{equation}
E_0 > E_g + E_2.
\end{equation}
If $M_c=0$, this condition can be met only if 
$a > 0.5 R_1 (R_\ast/R_1)^{1/2}$.
For $R_\ast/R_1 \sim 500$, this condition reads $a>10 R_1$.
However, we do expect the destruction to occur when $a <2 R_1$.
We conclude that a large fraction of the destructed companion
in our scenario must be accreted as a low entropy mass into the
companion.

 As an example, consider the case of $M_c = k M_2$, hence 
$M_{\rm acc}=(1-k) M_2$, $r_f  \simeq 0.25 R_1$, 
as the size of the solar core,
$R_\ast > 300 R_1$, and $a=2 R_1$. 
We also take the destruction energy of the companion to be 
\begin{equation}
E_2 = \beta \frac {G M_c M_{\rm acc}} {R_2}.
\end{equation}
The envelope of the companion before destruction, of mass
$M_{\rm acc}$, has a smaller average radius than $R_2$.
On the other hand, there is thermal energy of the companion.
Hence $\beta \sim 2-3$. 
Another energy source not considered here is the rotation energy
of each of the two stars, which are likely to corotate in our
scenario (see next section).  
Condition (7) becomes then
\begin{equation}
1 < 16 k - 4 \beta k (1-k) \frac{R_1 M_2}{R_2 M_1} 
+4 (1-k) \left(1+ k \frac{M_2}{M_1}\right) 
\left(\frac {R_1}{R_\ast} \right)^{1/2} .
\end{equation} 
 From this condition it is clearly seen that when $k=0$, the 
envelope can be inflated only to $R_\ast < 16 R_1$. 
For $k \ga 0.3$ and $R_\ast \gg R_1$, the last term, 
which is the energy of the inflated envelope can be neglected. 
Taking also $R_1 M_2 \simeq R_2 M_1$, the last condition
is simplified to 
\begin{equation}
1 < 4 k [4  - \beta (1-k)] 
\end{equation} 
for the assumed parameters here, the last condition is 
met by a large margin for $ k \ga 0.3$. 
This means that even by relaxing some of the assumptions used 
here, a giant may still be formed from such a process. 
 
Not all the energy will be retain by the accreted matter 
which build the envelope in our model. 
Some fraction of the energy goes to the fast wind and radiation as 
observed in the eruptions.
It is interesting to note that the radiation requires 
only small amount of mass to be accreted on the surface of the
original primary star.
Taking the energy of the effective-accreted mass that 
goes to radiation to be that of a Keplerian disk, 
we find this effective mass accretion rate to be
\begin{equation}
\dot M_{\rm rad} \simeq 6 \times 10^{-4} 
\left( \frac{L}{10^4 L_\odot} \right)
\left( \frac{R_1}{R_\odot} \right) 
\left( \frac{M_\ast}{M_\odot} \right)^{-1} M_\odot \mathrm{yr^{-1}} .
\end{equation}
For a distance to V838 Mon of $D =3$~kpc, the luminosity is 
$L \simeq 10^5 \mbox{L}_\odot$, and the accreted mass over several
years is still very small. 
 
\section{The proposed scenario}

Based on the previous section, in particular on our estimate that
a mass of $M_{\rm acc} \sim 0.05-0.3$ is required to inflate
the envelope to $\sim 500-1500 \mbox{R}_\odot$, we proposed the following general
scenario for the eruption of V838 Mon.
The progenitor was a close binary system composed of two 
main sequence stars: a primary with $M_1 \sim 1.5  M_\odot$,
and a companion with a mass 
$0.1 \mbox{M}_\odot \la \mbox{M}_2 \la 0.5 \mbox{M}_\odot$.
The small orbital separation implies that the spins of
both stars were synchronized with the orbital period, and the
orbit was circular.
Due to evolution, mainly magnetic winds, possibly from both stars,
the system lost angular momentum at a relatively fast rate, and the
orbital separation decreased.
One of the stars filled its Roche lobe, and transferred mass to
the other. 
For main sequence stars, the increase of stellar radius with 
the stellar mass is steeper that the increase of the average
Roche lobe size with mass (e.g., see formula by Eggleton \cite{egg}), and the 
more massive primary star is likely to fill its Roche lobe first. 
At about the same time as Roche lobe overflow occurs, a binary system
of two low mass main sequence stars becomes dynamically unstable 
(Rasio \& Shapiro \cite{rs}). 
This implies a very fast merger and mass transfer of the outer layers
of the stars. 
It is possible that the mass transfer of the outer layer of
the primary star onto the secondary caused the envelope to
be first inflated around the {\it secondary} star. 
In any case, mass transfer of the outer layers of the primary or
of both stars, we propose, caused the first burst in luminosity,
when V838 Mon reached a radius of $\sim 50 R_\odot$. 
Then, the secondary was accreted onto the primary, in particular the 
core of the secondary merged with the core of the primary, a process 
that released much more energy.
We proposed that this caused the second burst after about a month.

The merging process is likely to be complicated, with different epochs 
of variation in the luminosity, as parcells of gas being accreted and 
the merged stars, both envelope and the merging cores, are settled to a 
new equilibrium.
As noted in equation (11), a very small amount of accreted mass can 
lead to a relatively strong burst. 
Such a variation could cause the third outburst, which occurred about a month after
the second burst, and which required, if lasted one month,
only $\sim 5 \times 10^{-5} M_\odot (D/1\ \mathrm{kpc})^2$ to be accreted
on the initial surface of the primary star.

\section{Discussion and Summary}

Our main goal is to point out that the energy source of the outburst 
of V838 Mon was, and still is, an accretion event,
i.e., gravitational energy, rather than thermonuclear runaway.
We considered the merger of two main sequence stars, of masses 
$M_1 \simeq 1.5 M_\odot$, and $M_2 \simeq 0.1-0.5 M_\odot$.
In our proposed scenario, the first outburst, with a luminosity of 
$L_\ast \simeq 500 (D/1 {\rm kpc})^2 \mbox{L}_\odot$ and an envelope 
radius of $R_\ast \simeq 55 (D/1 {\rm kpc}) R_\odot$, was caused by 
the merger of the outer layers of the two stars, via dynamical instability. 
First mass was transfered from the more massive primary to the secondary;
most likely the envelope was first inflated around the secondary. 
Then, after about a months, the cores of the two stars collided,
releasing more energy, this led to the expansion of V838 Mon to 
$R_\ast \simeq 400 (D/1 {\rm kpc}) R_\odot$, and its luminosity to reache 
$L_\ast \simeq 1.0 \times 10^4 (D/1 {\rm kpc})^2 L_\odot$. 
Equation (11) shows that a small amount of accreted mass
can lead to large luminoisty variations. 
As the two merged stars settled to equilibrium, the luminoisty varies;
such a process caused the third outburst as well.

As mentioned in Sect. 1 a light echo have been detected around V838~Mon
during outburst. \cite{mhk} and Kimeswenger et~al. (\cite{kls}) interpret this as
due to scattering on circumstellar dust. This obviously rises 
a question about the origin of the circumstellar dust. In our scenario the
dust would reside in remains of a protostellar cloud from which the V838~Mon
binary system has been formed. This would imply a relatively young age of
V838~Mon which is consistent with its low distance ($\sim$13~pc) from 
the Galactic plane (\cite{mhk}). It is, however, also possible that the echo is
due to interestellar dust, as suggested by Bond et~al. (\cite{bps}), and then
obviously it has nothing to do with the nature of V838~Mon.

In principle, mergers of other types of stars is possible.
We note that the possible locations of the progenitor of V838 Mon
on the HR-diagram (\cite{mhk}), intersects the location of a zero-age
horizontal branch (HB) star which lost most of its envelope on the RGB 
(D'Cruz et al. \cite{cdr}). 
Merger of such an HB star with a low mass main sequence
star may leed to an inflated envelope around the HB star,
similar in many respects to the scenario proposed in the previous section.
The HB star lost most of its envelope because of the interaction with the 
low mass main sequence companion via a common envelope evolution,
which started when the HB progenitor was on the upper red ginat branch.
In this scenario the dust-halo can be attributed
to the high mass loss rate on the red giant branch.

Finally it is worth of noting that the effective photosphere of the three
objects seems to increase more or less steadilly with time during outburst.
In the case of M31~RV it expanded
up to $8000 {\rm R}_\odot$ (Mould et~al. \cite{mcg}). This would imply that
the observed photospheric regions are not in hydrostatic equillibrium. The
models of \cite{ht} and \cite{pl}, on which our considerations in Sect.~2 are based,
are in hydrostatic equillibrium. Slowly expanding envelopes would probably
require less mass and thus lower $M_{\rm acc}$ than hydrostatic
envelopes inflated to large radii. Thus accretion of an object of even lower
mass then estimated in Sect.~2 could give origin to the observed outbursts.

\begin{acknowledgements}
This work has partly been supported by the Polish State Committee for
Scientific Research through the grant No. 2.P03D.020.17. 
N. S. has also been supported 
by a grant from the USA-Israel Binational Science 
Foundation and the Israel Science Foundation.
\end{acknowledgements}

\end{document}